\documentclass[reprint, superscriptaddress]{revtex4-1}
\setcitestyle{super}

\usepackage{amsmath}

 \usepackage{ulem} 
 \usepackage{color}

\usepackage{graphicx}
\usepackage{natbib}

\usepackage[colorlinks=true,linkcolor=blue]{hyperref}%
\expandafter\ifx\csname package@font\endcsname\relax\else
 \expandafter\expandafter
 \expandafter\usepackage
 \expandafter\expandafter
 \expandafter{\csname package@font\endcsname}%
\fi
\makeindex
\begin{document}

\title{Tuning metal-insulator transitions in epitaxial V$_2$O$_3$ thin films }

\author{Einar B. Thorsteinsson}
\affiliation{Science Institute, University of Iceland, Dunhaga 3, 107 Reykjavik, Iceland}
\author{Seyedmohammad Shayestehaminzadeh}
\affiliation{Technovation Centre, AGC Glass Europe, Rue Louis Bl\'eriot 12, BE-6041 Gosselies, Belgium}
\author{Unnar B. Arnalds}
\affiliation{Science Institute, University of Iceland, Dunhaga 3, 107 Reykjavik, Iceland}
\email{uarnalds@hi.is}

\revised{\today}%

\begin{abstract}

We present a study of the synthesis of epitaxial V$_2$O$_3$ films on $c$-plane Al$_2$O$_3$ substrates by reactive dc-magnetron sputtering. The results reveal a temperature window, at substantially lower values than previously reported, wherein epitaxial films can be obtained when deposited on [0001] oriented surfaces. The films display a metal-insulator transition with a change in resistance of up to four orders of magnitude, strongly dependent on the O$_2$ partial pressure during deposition. While the electronic properties of the films show sensitivity to the amount of O$_2$ present during deposition of the films, their crystallographic structure and surface morphology of atomically flat terraced structures with up to micrometer dimensions are maintained. The transition temperature, as well as the scale of the metal-insulator transition, is correlated to the stoichiometry and local strain in the films controllable by the deposition parameters. 
\end{abstract}


\maketitle


Vanadium sesquioxide, V$_2$O$_3$, is a transition metal oxide which undergoes a first order structural phase transition (SPT) from a high temperature metallic state to a low temperature insulating state. During this structural transition the crystal structure of the film changes from a rhombohedral phase to a monoclinic phase and has been reported to exhibit a nanotextured phase coexistence in thin film form.\cite{McLeod_NPHYS:2017} The structural transition is linked to the metal-insulator transition (MIT) and can be observed through the resistivity of the material. For bulk V$_2$O$_3$, the transition temperature is around 155~K. However, for V$_2$O$_3$ thin films grown on single crystal substrates the temperature and magnitude of the MIT can vary with the choice of deposition method and conditions, substrate material and surface orientation.\cite{Sakai_2015, Brockman_nnano_2014, Valmianski_PRB_2017} The controlling materials parameters involved in these cases include strain in the film, induced by the lattice mismatch between V$_2$O$_3$ and substrate material,\cite{Schuler_TSF1997} the amount of vanadium and oxygen deficiencies,\cite{Brockman.APL.2011} crystalline defects, and structural disorder.\cite{Brockman.APL.2012}
Tunability of the metal-insulator transition by stoichiometry in bulk V$_{2-x}$O$_3$ has been observed where the increase in oxygen content shifts the MIT to lower temperatures and fully suppresses the transition for $x > 0.026$.\cite{UEDA_1980} 
A similar suppression of the MIT has been reported in Cr doped V$_2$O$_3$ thin films grown on Al$_2$O$_3$ [0001] surfaces and attributed to Cr promoting oxygen excess in the films, stabilizing the metallic state.\cite{Homm_2015} 

In this paper we present a study of the structural and electronic properties of V$_2$O$_3$ thin films fabricated on Al$_2$O$_3$ [0001] substrates by reactive dc-magnetron sputtering.
Controlling the oxygen partial pressure and substrate temperature during growth, we obtain epitaxial thin film layers of V$_2$O$_3$ showing low roughness terraced surfaces. The structural quality of the films is linked to the substrate temperature during deposition and resilient to changes in the O$_2$ partial pressure during deposition. However, we observe a strong dependence of the magnitude and temperature of the MIT on the O$_2$ partial pressure during growth. 


The films were fabricated by reactive dc-magnetron sputtering from a vanadium target using a custom built magnetron sputtering chamber\cite{Arnalds_07_RSI}. During sputtering the chamber pressure was maintained at 0.4 Pa with an Ar flow rate of 20 sccm and O$_2$ flow rate in the range 1.4 to 2.0 sccm. The films were deposited onto unannealed single crystalline sapphire substrates with $c$-plane [0001] surface orientations. During growth, the substrate temperature was controlled in the range of 350$^\circ$C to 670$^\circ$C. After film growth the samples were allowed to cool down {\it in-situ} to room temperature before {\it ex-situ} characterization. The thickness of the films was maintained at $\sim$60~nm. The deposition rate varied slightly for the different O$_2$ settings rising from 0.84 {\AA}/s at 1.4 sccm up to a maximum of 0.89 \AA/s at 1.6 sccm before falling to 0.77 \AA/s at 2.0 sccm. The structural properties and surface morphology of the films were investigated by x-ray diffraction, x-ray reflectivity measurements, reciprocal space mapping and atomic force microscopy. The electrical resistance of the films was measured using a two point resistance measurement setup in the temperature range of 10~K to 300~K. 100 nm thick gold electrodes were deposited along the edges of the films after deposition to provide electrical contact to the films. Further details regarding the experimental procedures are given in supplementary material.



\begin{figure}[t]
  \begin{center}
   \includegraphics[width=\columnwidth]{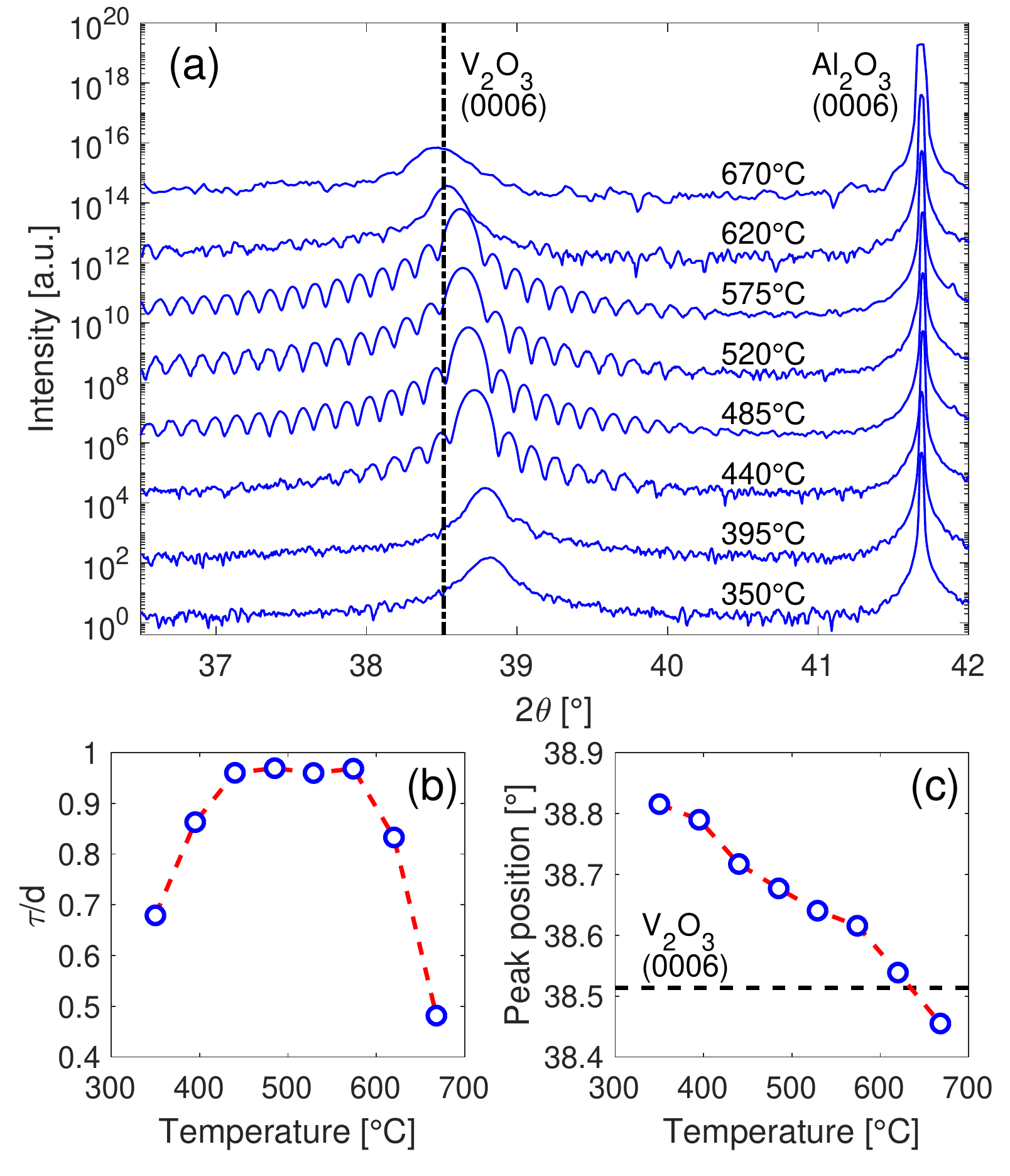} 
  \end{center}
  \caption{Dependence of structural properties on temperature. (a) X-ray diffraction scans recorded for films deposited at different substrate temperatures for a fixed O$_2$ flow rate of 1.6 sccm, Ar flow rate of 20 sccm, chamber pressure of 0.4~Pa and constant sputtering power of 150 W.  (b) Ratio of the mean size of the crystalline domains in the vertical direction, $\tau$, to film thickness as a function of substrate temperature during deposition. The perpendicular coherence, $\tau$, was determined from the FWHM of the V$_2$O$_3$ peak using the Scherrer formula. (c) V$_2$O$_3$ (0006) angular peak position. The dashed lines in (a) and (c) indicate the angular peak position of bulk V$_2$O$_3$. 
    }
     \label{fig:xray_temp}
\end{figure}

Figure \ref{fig:xray_temp} shows x-ray diffraction scans recorded for films grown onto $c$-plane Al$_2$O$_3$ for various substrate temperatures while maintaining a fixed O$_2$ partial pressure. Varying the growth temperature reveals a finite temperature window within 400$^\circ$C to 600$^\circ$C (for the chosen O$_2$ flow rate).  Within this range, the films display a highly epitaxial nature with a strong V$_2$O$_3$ (0006) peak and Laue fringes extending from both sides of the peak. This is highlighted in Figure \ref{fig:xray_temp}(b) which shows the ratio of the mean size of crystalline domains in the vertical direction to film thickness, as a function of temperature. 
Previously reported temperature values used for the fabrication of epitaxial V$_2$O$_3$ are generally in the range of $700^{\circ}$C.\cite{Dillemans_APL_2014, Gilbert_2017, Venta.APL.2014, Brockman.APL.2012, Brockman_nnano_2014, Sass_2004} Our results however show that epitaxial V$_2$O$_3$ films can be obtained using reactive dc-magnetron sputtering at deposition temperature values substantially lower than previously reported and without the need for post deposition annealing.\cite{Ji.APL.2012}
Figure \ref{fig:xray_temp}(c) shows the location of the V$_2$O$_3$ (0006) peak shifting toward lower values with increasing temperature crossing the bulk value of 38.514$^\circ$.\cite{NBS_V2O3_bulk} The shift of the peak to higher angles compared to bulk V$_2$O$_3$ indicates compressive strain in the out-of-plane direction relaxing with increasing deposition temperature. 
 
\begin{figure}[t]
  \begin{center}
   \includegraphics[width=\columnwidth]{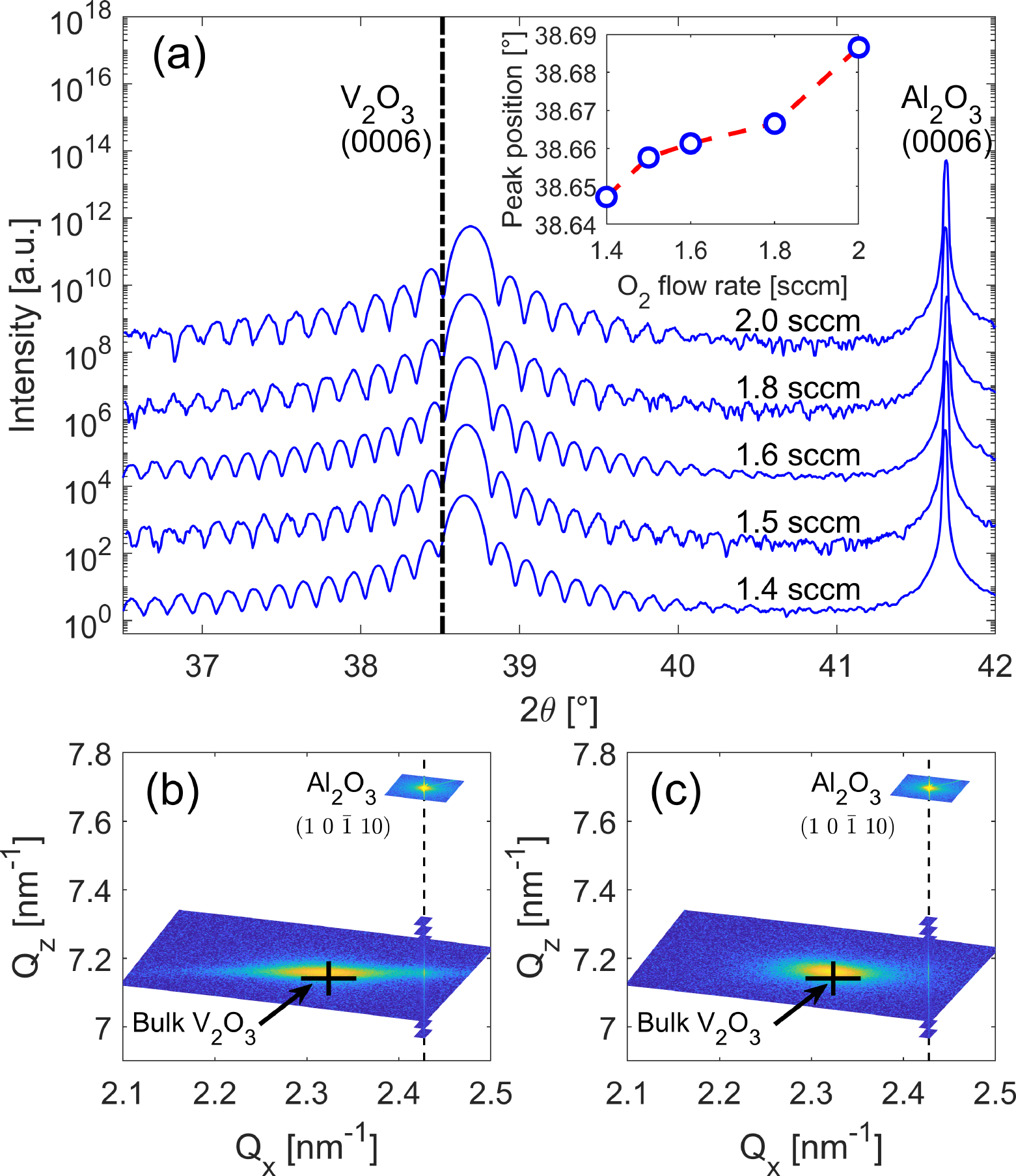} \\
   \end{center}
  \caption{Structural resilience with regards to O$_2$ conditions. (a) X-ray diffraction scans for different O$_2$ flow rates with other deposition parameters fixed (substrate temperature of 485$^{\circ}$C and sputtering power of 150 W). The inset shows the position of the V$_2$O$_3$ peak for the different O$_2$ conditions. Graphs (b) and (c) show reciprocal space maps for films grown with 1.4 sccm and 2.0 sccm flow rates, respectively, around the (1 0 -1 10) peaks of V$_2$O$_3$ and Al$_2$O$_3$. RSM scans for the series is provided in supplementary material.}
     \label{fig:xray_O2}
\end{figure}

Selecting a fixed substrate temperature of 485$^{\circ}$C, from the x-ray investigations shown in Figure \ref{fig:xray_temp}, a series of films were grown at different O$_2$ flow rate settings while maintaining other growth parameters fixed. Figure \ref{fig:xray_O2}(a) shows x-ray diffraction scans of a series of films deposited at different O$_2$ flow rates from 1.4 sccm to 2.0 sccm. For these films, only slight shifts of the peak positions (to within 0.04$^{\circ}$) are observed for the different O$_2$ deposition conditions. The corresponding values of the $c$-lattice parameter range between 13.967~\AA\ and 13.954~\AA. In all cases the films consistently show well-defined Laue oscillations and reveal a vertical coherence close to the film thickness. Full width at half maximum (FWHM) of rocking curves ($\omega$ scans) recorded around the (0006) peaks are in all cases below 0.003$^{\circ}$. The epitaxial nature of the films is therefore resilient to variations within this O$_2$ flow rate range. 

\begin{figure}[t]
  \begin{center}
   \includegraphics[width=\columnwidth]{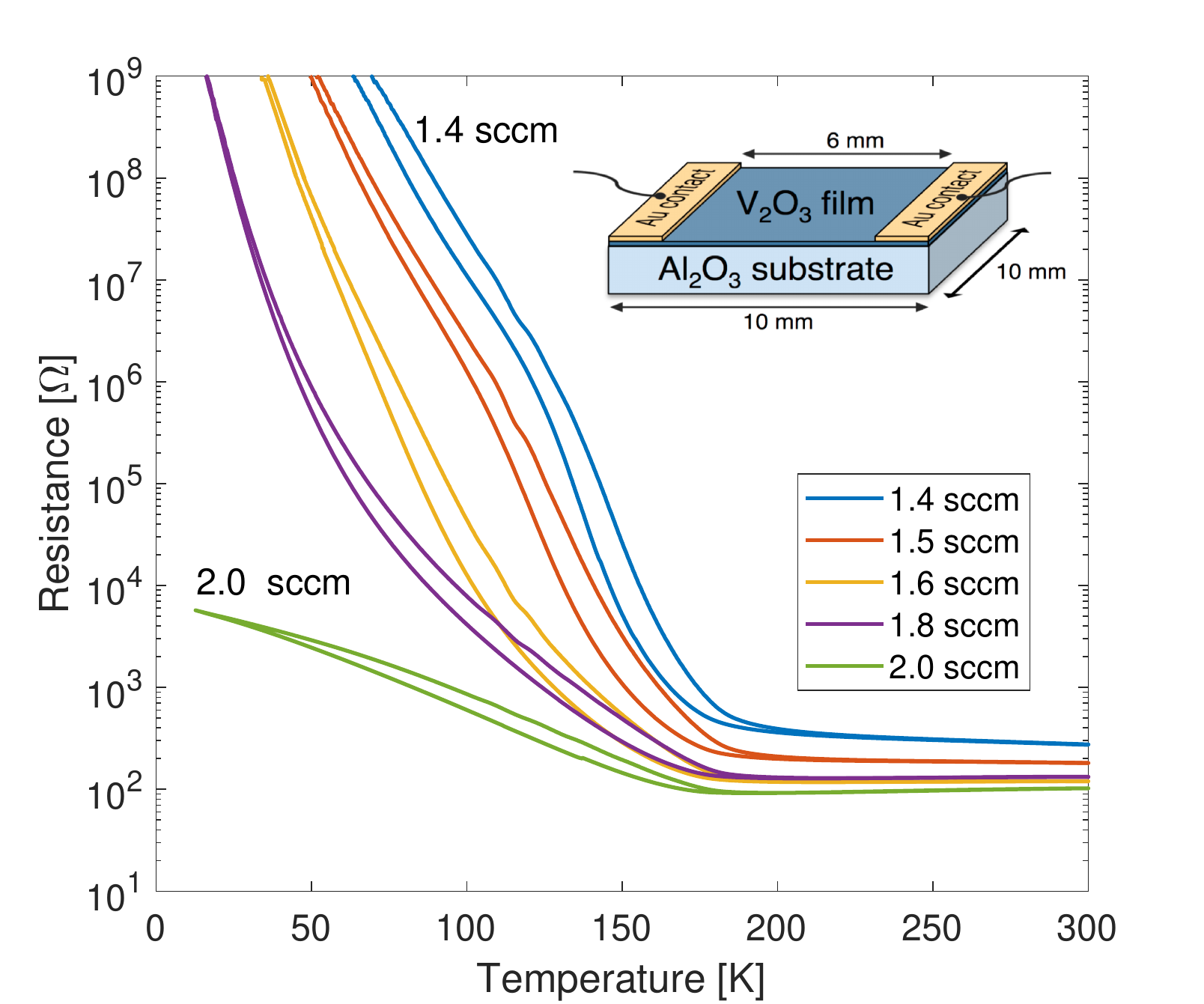} 
  \end{center}
  \caption{Resistance as a function of temperature for the $\sim$60~nm thick V$_2$O$_3$ films deposited under different oxygen conditions. The inset schematic illustrates the sample structure where the resistance of the films was measured using 100~nm thick and 2~mm wide gold contacts deposited along the edges of the films.}
     \label{fig:O2_resistance}
\end{figure}

Figure \ref{fig:O2_resistance} shows the resistance as a function of temperature for films fabricated under the different O$_2$ conditions. All the films display a clear MIT with varying magnitude. For these films the resistance change ranges up to four orders of magnitude for the transition. As the temperature is further reduced the resistance increases resulting in a total resistance change of up to 7 orders of magnitude, limited by the measurement setup. As illustrated in Figure \ref{fig:xray_O2} the structural properties of the films are preserved. However, under the variation of the oxygen conditions the temperature dependent resistance of the films changes. For the films fabricated under the lower oxygen flows, a clear abrupt transition is observed with a transition temperature close to that of bulk V$_2$O$_3$. As the oxygen flow rate increases, both the magnitude and temperature of the transition are reduced and finally suppressed.
Further reduction of O$_2$ flow settings (with other growth parameters fixed) was not performed for this study. However, increasing the sputtering power revealed a further shift towards higher temperatures (see supplementary material) beyond the highest temperature transition recorded and shown in Fig. \ref{fig:O2_resistance}. For these samples the steepness of the transition was reduced in agreement with observations by Brockman et. al for films grown with molecular beam epitaxy.\cite{Brockman.APL.2011} 

In order to further investigate the epitaxial nature of the films and explore the strain within the grown films, we performed reciprocal space map (RSM) scans of the films around the asymmetrical (1 0 -1 10) peak. The (1 0 -1 10) peak was chosen as it has the strongest relative intensity for V$_2$O$_3$.  For all the films, the scans showed peaks confirming the epitaxy of the films. Figures \ref{fig:xray_O2}(b) and (c) show RSM scans for films grown at 1.4 sccm and 2.0 sccm O$_2$ flow rates. The scans show two layer peaks, one corresponding to a fully strained component of the film and one corresponding to a relaxed part with position close to bulk values. The extracted $c$ lattice parameters of the relaxed component show a slight shift towards lower lattice spacing consistent with the shift in the (0006) peak in the x-ray diffraction scans (Fig. \ref{fig:xray_O2}(a)). Extracting the lateral $a$ lattice parameter gives values in the range of $a = 4.977$~\AA\  to $a = 4.971$~\AA\ for the relaxed part of the film from lowest to highest O$_2$ flow settings. Compared to bulk V$_2$O$_3$ with $a = 4.954$~\AA\ these values show the lattice to be expanded in the plane of the film. Considering the in-plane lattice parameter of Al$_2$O$_3$, which is 4.1\% smaller than that of bulk V$_2$O$_3$, this result runs counter to expectation as the relaxed V$_2$O$_3$ part is expanded in the plane, increasing the lattice mismatch.\cite{Schuler_TSF1997} The variation in the $c$ and $a$ lattice parameters we observe for the different O$_2$ conditions are an order of magnitude lower than that expected for thermal expansion due to the temperature during growth. Fitting the peak widths in Figures \ref{fig:xray_O2} (b) and (c) reveals a mosaic spread in the range $0.11^{\circ}$ to $0.28^{\circ}$ and a lateral correlation length ranging from 15 nm to 55 nm with increasing O$_2$ flow (see supplementary material).  \cite{Eckert_JAP_1973, Munro_1997} The peaks with $Q_x$ values corresponding to the substrate indicate that a fully strained interface layer is formed in the initial stages of deposition.\cite{Dillemans_APL_2014, Brockman.APL.2012} The intensity of these peaks reduces with increased O$_2$ flow rates.


The observed results support the structural model proposed by Schuler {\it et al.}\cite{Schuler_TSF1997} concerning the local stress evolution as the film grows. Although the strain-relaxation in the higher O-containing film is more pronounced, the local tensile strain in the vicinity of the film surface is higher than that of the low O-containing film due to more inhabitation of O interstitials in the crystal. This is evident considering the lateral structural correlation length being larger as a result of larger lattice mismatch which causes long-range elastic interaction between different domains of the film, specifically during the phase transition.\cite{McLeod_NPHYS:2017}  The oxygen excess we observe in our films is corroborated by the lattice parameters determined from the x-ray scans. For bulk V$_{2-x}$O$_{3}$, suppression of the MIT has been observed for $x>0.026$.\cite{UEDA_1980} The concomitant variations in the $a$ and $c$ lattice parameters reported, (4.947~\AA $<$a$<$4.955~\AA) and (13.995~\AA$<$c$<$14.003~\AA), correspond closely to the values we observe for our thin film variant.

As recently reported \cite{McLeod_NPHYS:2017}, the phase transition in V$_2$O$_3$ undergoes nanotexturing of insulating and metallic domains where the domains co-existence \cite{Gao_2002, Bratkovsky_1994} and dynamics are governed by the lattice-mismatch-induced elastic interaction between monoclinic and rhombohedral domains. However, during the transition, the dynamics of the emerging insulating domain can be suppressed by the stabilization of metallic domains by local stress in the film.\cite{McLeod_NPHYS:2017} Such stress can induce alteration in $d$-orbital occupancy \cite{Tokura_Science_2000} inducing a collapse in insulating domain emergence. This phenomenon has been observed for VO$_2$ thin film systems under photoexcitation,\cite{He_PRB_2016, Morrison_Science_2014} under controlled epitaxial strain,\cite{Aetukuri_NatPhys_2013} predicted for other correlated metal oxide systems,\cite{Verma_2016} and supported by nano-IR imaging for V$_2$O$_3$.\cite{McLeod_NPHYS:2017} 
Although metal domain stabilization has been speculated to be due to epitaxial strain, our results indicate that the local stress evolution is responsible for the change in overall film conductivity. The stress evolution in our films is mediated by higher oxygen content, which we relate to more O-interstitials promoting elevated mismatch and local tensile stress in the vicinity of the surface. 

\begin{figure}[t]
  \begin{center}
   \includegraphics[width=0.98\columnwidth]{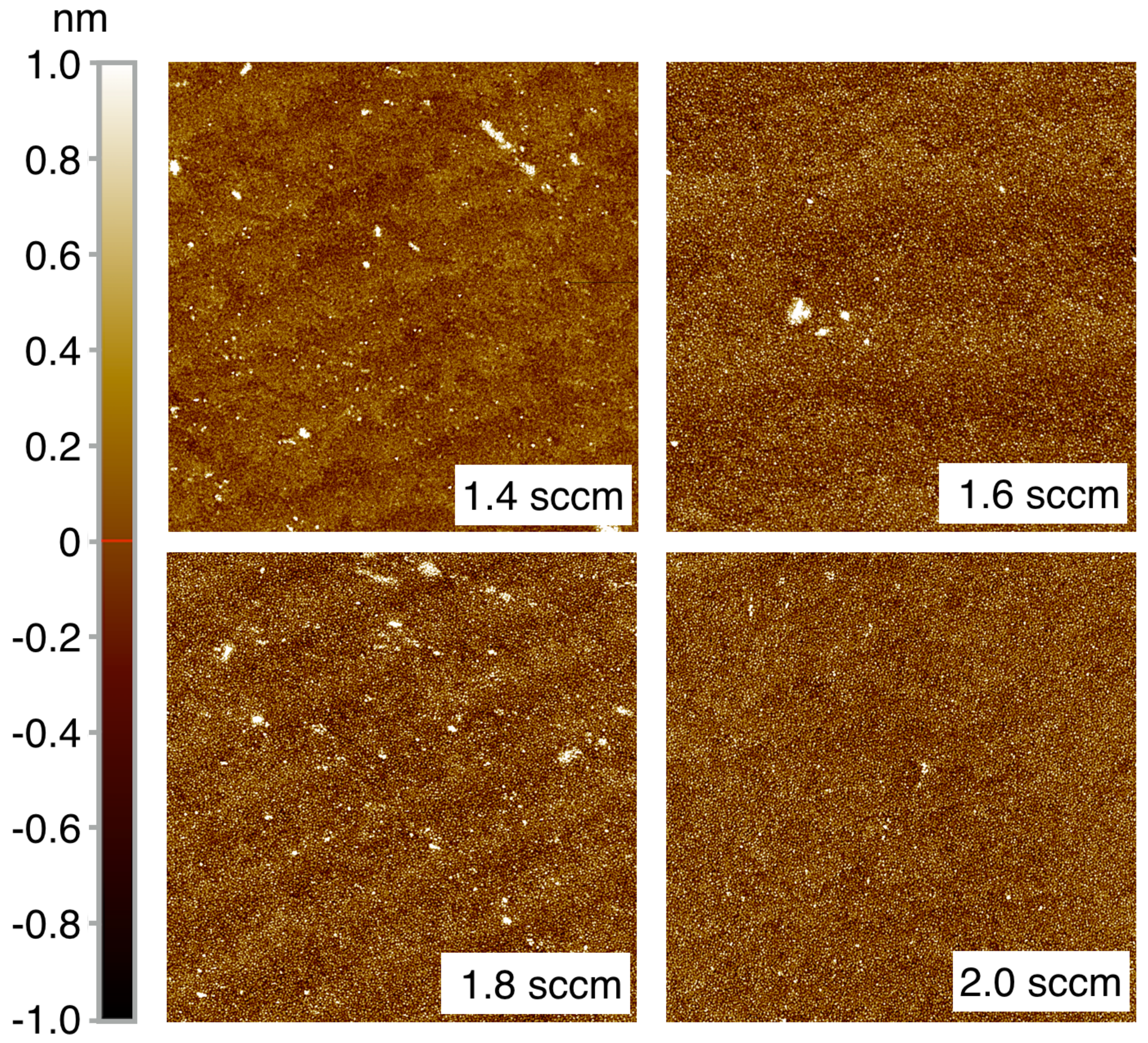} 
  \end{center}
  \caption{$5\times5$~$\mu$m$^2$ atomic force microscopy images of V$_2$O$_3$ films deposited under varying O$_2$ flow rates with a fixed substrate temperature of 485$^{\circ}$C.  The RMS roughness values for the images are 0.22 nm, 0.37 nm, 0.40 nm, and 0.40 nm, in the order of increasing O$_2$ flow. }
     \label{fig:afm}
\end{figure}

Figure \ref{fig:afm} shows AFM images of the sample series where the O$_2$ partial pressure was varied (see Figure \ref{fig:xray_O2}). With varying O$_2$ flow rates, the surfaces show large terraced structures indicating atomically flat epitaxial films in agreement with the low mosaic spread observed from the RSM. 
AFM images recorded for films grown at temperatures at the lower and higher end of the temperature window (see figure \ref{fig:xray_temp}) showed an increased surface roughness. 
Gilbert {\it et al.} have recently observed an induced uniaxial anisotropy in Ni films deposited on underlying V$_2$O$_3$ layers.\cite{Gilbert_2017} This induced anisotropy is attributed to steps in the Al$_2$O$_3$ substrates inducing rips in the overlying Ni layer affecting the magnetization reversal via domain wall pinning. Here, we show that atomic terraces can be obtained directly for sputter deposited V$_2$O$_3$ films on unannealed $c$-plane Al$_2$O$_3$ surfaces. Proximity effects between V$_2$O$_3$ and overlying magnetic layers in hybrid thin film heterostructures could therefore be tuned by the morphology at the atomic level of the V$_2$O$_3$ surface. \cite{Venta.APL.2014}


In summary, by controlling the deposition conditions for reactively sputtered V$_2$O$_3$ films we have fabricated flat epitaxial films with surface roughness below 0.5~nm at deposition temperatures below 500$^{\circ}$C. Tuning the O$_2$ partial pressure, we are able to control the temperature and scale of the metal-insulator transition through stoichiometry and local strain engineering of the films, altering orbital occupancy and metallicity stabilization while maintaining the crystal structure minorly affected. 

\subsection{Supplementary material}

See supplementary material for further details regarding thin film growth, characterization methods, reciprocal space maps of the sample sets presented in Figure \ref{fig:xray_O2} and resistivity results on complimentary samples.

\subsection{Acknowledgments}

This work was supported by funding from the Icelandic Research Fund Grant No. 141518-051 and Grant No. 152483-051 and the University of Iceland Research fund. The authors acknowledge discussions with Arni S. Ingason.


\begin{thebibliography}{25}%
\makeatletter
\providecommand \@ifxundefined [1]{%
 \@ifx{#1\undefined}
}%
\providecommand \@ifnum [1]{%
 \ifnum #1\expandafter \@firstoftwo
 \else \expandafter \@secondoftwo
 \fi
}%
\providecommand \@ifx [1]{%
 \ifx #1\expandafter \@firstoftwo
 \else \expandafter \@secondoftwo
 \fi
}%
\providecommand \natexlab [1]{#1}%
\providecommand \enquote  [1]{``#1''}%
\providecommand \bibnamefont  [1]{#1}%
\providecommand \bibfnamefont [1]{#1}%
\providecommand \citenamefont [1]{#1}%
\providecommand \href@noop [0]{\@secondoftwo}%
\providecommand \href [0]{\begingroup \@sanitize@url \@href}%
\providecommand \@href[1]{\@@startlink{#1}\@@href}%
\providecommand \@@href[1]{\endgroup#1\@@endlink}%
\providecommand \@sanitize@url [0]{\catcode `\\12\catcode `\$12\catcode
  `\&12\catcode `\#12\catcode `\^12\catcode `\_12\catcode `\%12\relax}%
\providecommand \@@startlink[1]{}%
\providecommand \@@endlink[0]{}%
\providecommand \url  [0]{\begingroup\@sanitize@url \@url }%
\providecommand \@url [1]{\endgroup\@href {#1}{\urlprefix }}%
\providecommand \urlprefix  [0]{URL }%
\providecommand \Eprint [0]{\href }%
\providecommand \doibase [0]{http://dx.doi.org/}%
\providecommand \selectlanguage [0]{\@gobble}%
\providecommand \bibinfo  [0]{\@secondoftwo}%
\providecommand \bibfield  [0]{\@secondoftwo}%
\providecommand \translation [1]{[#1]}%
\providecommand \BibitemOpen [0]{}%
\providecommand \bibitemStop [0]{}%
\providecommand \bibitemNoStop [0]{.\EOS\space}%
\providecommand \EOS [0]{\spacefactor3000\relax}%
\providecommand \BibitemShut  [1]{\csname bibitem#1\endcsname}%
\let\auto@bib@innerbib\@empty
\bibitem [{\citenamefont {McLeod}\ \emph {et~al.}(2017)\citenamefont {McLeod},
  \citenamefont {van Heumen}, \citenamefont {Ramirez}, \citenamefont {Wang},
  \citenamefont {Saerbeck}, \citenamefont {Guenon}, \citenamefont {Goldflam},
  \citenamefont {Anderegg}, \citenamefont {Kelly}, \citenamefont {Mueller},
  \citenamefont {Liu}, \citenamefont {Schuller},\ and\ \citenamefont
  {Basov}}]{McLeod_NPHYS:2017}%
  \BibitemOpen
  \bibfield  {author} {\bibinfo {author} {\bibfnamefont {A.~S.}\ \bibnamefont
  {McLeod}}, \bibinfo {author} {\bibfnamefont {E.}~\bibnamefont {van Heumen}},
  \bibinfo {author} {\bibfnamefont {J.~G.}\ \bibnamefont {Ramirez}}, \bibinfo
  {author} {\bibfnamefont {S.}~\bibnamefont {Wang}}, \bibinfo {author}
  {\bibfnamefont {T.}~\bibnamefont {Saerbeck}}, \bibinfo {author}
  {\bibfnamefont {S.}~\bibnamefont {Guenon}}, \bibinfo {author} {\bibfnamefont
  {M.}~\bibnamefont {Goldflam}}, \bibinfo {author} {\bibfnamefont
  {L.}~\bibnamefont {Anderegg}}, \bibinfo {author} {\bibfnamefont
  {P.}~\bibnamefont {Kelly}}, \bibinfo {author} {\bibfnamefont
  {A.}~\bibnamefont {Mueller}}, \bibinfo {author} {\bibfnamefont {M.~K.}\
  \bibnamefont {Liu}}, \bibinfo {author} {\bibfnamefont {I.~K.}\ \bibnamefont
  {Schuller}}, \ and\ \bibinfo {author} {\bibfnamefont {D.~N.}\ \bibnamefont
  {Basov}},\ }\href {http://dx.doi.org/10.1038/nphys3882} {\bibfield  {journal}
  {\bibinfo  {journal} {Nat Phys}\ }\textbf {\bibinfo {volume} {13}},\ \bibinfo
  {pages} {80} (\bibinfo {year} {2017})}\BibitemShut {NoStop}%
\bibitem [{\citenamefont {Sakai}\ \emph {et~al.}(2015)\citenamefont {Sakai},
  \citenamefont {Limelette},\ and\ \citenamefont {Funakubo}}]{Sakai_2015}%
  \BibitemOpen
  \bibfield  {author} {\bibinfo {author} {\bibfnamefont {J.}~\bibnamefont
  {Sakai}}, \bibinfo {author} {\bibfnamefont {P.}~\bibnamefont {Limelette}}, \
  and\ \bibinfo {author} {\bibfnamefont {H.}~\bibnamefont {Funakubo}},\ }\href
  {\doibase 10.1063/1.4937456} {\bibfield  {journal} {\bibinfo  {journal}
  {Applied Physics Letters}\ }\textbf {\bibinfo {volume} {107}},\ \bibinfo
  {pages} {241901} (\bibinfo {year} {2015})}\BibitemShut {NoStop}%
\bibitem [{\citenamefont {Brockman}\ \emph {et~al.}(2014)\citenamefont
  {Brockman}, \citenamefont {Gao}, \citenamefont {Hughes}, \citenamefont
  {Rettner}, \citenamefont {Samant}, \citenamefont {Roche},\ and\ \citenamefont
  {Parkin}}]{Brockman_nnano_2014}%
  \BibitemOpen
  \bibfield  {author} {\bibinfo {author} {\bibfnamefont {J.~S.}\ \bibnamefont
  {Brockman}}, \bibinfo {author} {\bibfnamefont {L.}~\bibnamefont {Gao}},
  \bibinfo {author} {\bibfnamefont {B.}~\bibnamefont {Hughes}}, \bibinfo
  {author} {\bibfnamefont {C.~T.}\ \bibnamefont {Rettner}}, \bibinfo {author}
  {\bibfnamefont {M.~G.}\ \bibnamefont {Samant}}, \bibinfo {author}
  {\bibfnamefont {K.~P.}\ \bibnamefont {Roche}}, \ and\ \bibinfo {author}
  {\bibfnamefont {S.~S.~P.}\ \bibnamefont {Parkin}},\ }\href
  {http://dx.doi.org/10.1038/nnano.2014.71} {\bibfield  {journal} {\bibinfo
  {journal} {Nat Nano}\ }\textbf {\bibinfo {volume} {9}},\ \bibinfo {pages}
  {453} (\bibinfo {year} {2014})}\BibitemShut {NoStop}%
\bibitem [{\citenamefont {Valmianski}\ \emph {et~al.}(2017)\citenamefont
  {Valmianski}, \citenamefont {Ramirez}, \citenamefont {Urban}, \citenamefont
  {Batlle},\ and\ \citenamefont {Schuller}}]{Valmianski_PRB_2017}%
  \BibitemOpen
  \bibfield  {author} {\bibinfo {author} {\bibfnamefont {I.}~\bibnamefont
  {Valmianski}}, \bibinfo {author} {\bibfnamefont {J.~G.}\ \bibnamefont
  {Ramirez}}, \bibinfo {author} {\bibfnamefont {C.}~\bibnamefont {Urban}},
  \bibinfo {author} {\bibfnamefont {X.}~\bibnamefont {Batlle}}, \ and\ \bibinfo
  {author} {\bibfnamefont {I.~K.}\ \bibnamefont {Schuller}},\ }\href {\doibase
  10.1103/PhysRevB.95.155132} {\bibfield  {journal} {\bibinfo  {journal} {Phys.
  Rev. B}\ }\textbf {\bibinfo {volume} {95}},\ \bibinfo {pages} {155132}
  (\bibinfo {year} {2017})}\BibitemShut {NoStop}%
\bibitem [{\citenamefont {Schuler}\ \emph {et~al.}(1997)\citenamefont
  {Schuler}, \citenamefont {Klimm}, \citenamefont {Weissmann}, \citenamefont
  {Renner},\ and\ \citenamefont {Horn}}]{Schuler_TSF1997}%
  \BibitemOpen
  \bibfield  {author} {\bibinfo {author} {\bibfnamefont {H.}~\bibnamefont
  {Schuler}}, \bibinfo {author} {\bibfnamefont {S.}~\bibnamefont {Klimm}},
  \bibinfo {author} {\bibfnamefont {G.}~\bibnamefont {Weissmann}}, \bibinfo
  {author} {\bibfnamefont {C.}~\bibnamefont {Renner}}, \ and\ \bibinfo {author}
  {\bibfnamefont {S.}~\bibnamefont {Horn}},\ }\href {\doibase
  https://doi.org/10.1016/S0040-6090(96)09399-6} {\bibfield  {journal}
  {\bibinfo  {journal} {Thin Solid Films}\ }\textbf {\bibinfo {volume} {299}},\
  \bibinfo {pages} {119 } (\bibinfo {year} {1997})}\BibitemShut {NoStop}%
\bibitem [{\citenamefont {Brockman}\ \emph {et~al.}(2011)\citenamefont
  {Brockman}, \citenamefont {Aetukuri}, \citenamefont {Topuria}, \citenamefont
  {Samant}, \citenamefont {Roche},\ and\ \citenamefont
  {Parkin}}]{Brockman.APL.2011}%
  \BibitemOpen
  \bibfield  {author} {\bibinfo {author} {\bibfnamefont {J.}~\bibnamefont
  {Brockman}}, \bibinfo {author} {\bibfnamefont {N.~P.}\ \bibnamefont
  {Aetukuri}}, \bibinfo {author} {\bibfnamefont {T.}~\bibnamefont {Topuria}},
  \bibinfo {author} {\bibfnamefont {M.~G.}\ \bibnamefont {Samant}}, \bibinfo
  {author} {\bibfnamefont {K.~P.}\ \bibnamefont {Roche}}, \ and\ \bibinfo
  {author} {\bibfnamefont {S.~S.~P.}\ \bibnamefont {Parkin}},\ }\href
  {https://doi.org/10.1063/1.3574910} {\bibfield  {journal} {\bibinfo
  {journal} {Applied Physics Letters}\ }\textbf {\bibinfo {volume} {98}},\
  \bibinfo {pages} {152105} (\bibinfo {year} {2011})}\BibitemShut {NoStop}%
\bibitem [{\citenamefont {Brockman}\ \emph {et~al.}(2012)\citenamefont
  {Brockman}, \citenamefont {Samant}, \citenamefont {Roche},\ and\
  \citenamefont {Parkin}}]{Brockman.APL.2012}%
  \BibitemOpen
  \bibfield  {author} {\bibinfo {author} {\bibfnamefont {J.}~\bibnamefont
  {Brockman}}, \bibinfo {author} {\bibfnamefont {M.~G.}\ \bibnamefont
  {Samant}}, \bibinfo {author} {\bibfnamefont {K.~P.}\ \bibnamefont {Roche}}, \
  and\ \bibinfo {author} {\bibfnamefont {S.~S.~P.}\ \bibnamefont {Parkin}},\
  }\href
  {http://scitation.aip.org/content/aip/journal/apl/101/5/10.1063/1.4742160}
  {\bibfield  {journal} {\bibinfo  {journal} {Applied Physics Letters}\
  }\textbf {\bibinfo {volume} {101}},\ \bibinfo {eid} {051606} (\bibinfo {year}
  {2012})}\BibitemShut {NoStop}%
\bibitem [{\citenamefont {Ueda}\ \emph {et~al.}(1980)\citenamefont {Ueda},
  \citenamefont {Kosuge},\ and\ \citenamefont {Kachi}}]{UEDA_1980}%
  \BibitemOpen
  \bibfield  {author} {\bibinfo {author} {\bibfnamefont {Y.}~\bibnamefont
  {Ueda}}, \bibinfo {author} {\bibfnamefont {K.}~\bibnamefont {Kosuge}}, \ and\
  \bibinfo {author} {\bibfnamefont {S.}~\bibnamefont {Kachi}},\ }\href@noop {}
  {\bibfield  {journal} {\bibinfo  {journal} {Journal of Solid State
  Chemistry}\ }\textbf {\bibinfo {volume} {31}},\ \bibinfo {pages} {171}
  (\bibinfo {year} {1980})}\BibitemShut {NoStop}%
\bibitem [{\citenamefont {Homm}\ \emph {et~al.}(2015)\citenamefont {Homm},
  \citenamefont {Dillemans}, \citenamefont {Menghini}, \citenamefont
  {Van~Bilzen}, \citenamefont {Bakalov}, \citenamefont {Su}, \citenamefont
  {Lieten}, \citenamefont {Houssa}, \citenamefont {Nasr~Esfahani},
  \citenamefont {Covaci}, \citenamefont {Peeters}, \citenamefont {Seo},\ and\
  \citenamefont {Locquet}}]{Homm_2015}%
  \BibitemOpen
  \bibfield  {author} {\bibinfo {author} {\bibfnamefont {P.}~\bibnamefont
  {Homm}}, \bibinfo {author} {\bibfnamefont {L.}~\bibnamefont {Dillemans}},
  \bibinfo {author} {\bibfnamefont {M.}~\bibnamefont {Menghini}}, \bibinfo
  {author} {\bibfnamefont {B.}~\bibnamefont {Van~Bilzen}}, \bibinfo {author}
  {\bibfnamefont {P.}~\bibnamefont {Bakalov}}, \bibinfo {author} {\bibfnamefont
  {C.-Y.}\ \bibnamefont {Su}}, \bibinfo {author} {\bibfnamefont
  {R.}~\bibnamefont {Lieten}}, \bibinfo {author} {\bibfnamefont
  {M.}~\bibnamefont {Houssa}}, \bibinfo {author} {\bibfnamefont
  {D.}~\bibnamefont {Nasr~Esfahani}}, \bibinfo {author} {\bibfnamefont
  {L.}~\bibnamefont {Covaci}}, \bibinfo {author} {\bibfnamefont {F.~M.}\
  \bibnamefont {Peeters}}, \bibinfo {author} {\bibfnamefont {J.~W.}\
  \bibnamefont {Seo}}, \ and\ \bibinfo {author} {\bibfnamefont {J.-P.}\
  \bibnamefont {Locquet}},\ }\href
  {http://scitation.aip.org/content/aip/journal/apl/107/11/10.1063/1.4931372}
  {\bibfield  {journal} {\bibinfo  {journal} {Applied Physics Letters}\
  }\textbf {\bibinfo {volume} {107}},\ \bibinfo {eid} {111904} (\bibinfo {year}
  {2015})}\BibitemShut {NoStop}%
\bibitem [{\citenamefont {Arnalds}\ \emph {et~al.}(2007)\citenamefont
  {Arnalds}, \citenamefont {Agustsson}, \citenamefont {Ingason}, \citenamefont
  {Eriksson}, \citenamefont {Gylfason}, \citenamefont {Gudmundsson},\ and\
  \citenamefont {Olafsson}}]{Arnalds_07_RSI}%
  \BibitemOpen
  \bibfield  {author} {\bibinfo {author} {\bibfnamefont {U.~B.}\ \bibnamefont
  {Arnalds}}, \bibinfo {author} {\bibfnamefont {J.~S.}\ \bibnamefont
  {Agustsson}}, \bibinfo {author} {\bibfnamefont {A.~S.}\ \bibnamefont
  {Ingason}}, \bibinfo {author} {\bibfnamefont {A.~K.}\ \bibnamefont
  {Eriksson}}, \bibinfo {author} {\bibfnamefont {K.~B.}\ \bibnamefont
  {Gylfason}}, \bibinfo {author} {\bibfnamefont {J.~T.}\ \bibnamefont
  {Gudmundsson}}, \ and\ \bibinfo {author} {\bibfnamefont {S.}~\bibnamefont
  {Olafsson}},\ }\href {\doibase 10.1063/1.2793508} {\bibfield  {journal}
  {\bibinfo  {journal} {Review of Scientific Instruments}\ }\textbf {\bibinfo
  {volume} {78}},\ \bibinfo {pages} {103901} (\bibinfo {year}
  {2007})}\BibitemShut {NoStop}%
\bibitem [{\citenamefont {Dillemans}\ \emph {et~al.}(2014)\citenamefont
  {Dillemans}, \citenamefont {Smets}, \citenamefont {Lieten}, \citenamefont
  {Menghini}, \citenamefont {Su},\ and\ \citenamefont
  {Locquet}}]{Dillemans_APL_2014}%
  \BibitemOpen
  \bibfield  {author} {\bibinfo {author} {\bibfnamefont {L.}~\bibnamefont
  {Dillemans}}, \bibinfo {author} {\bibfnamefont {T.}~\bibnamefont {Smets}},
  \bibinfo {author} {\bibfnamefont {R.~R.}\ \bibnamefont {Lieten}}, \bibinfo
  {author} {\bibfnamefont {M.}~\bibnamefont {Menghini}}, \bibinfo {author}
  {\bibfnamefont {C.-Y.}\ \bibnamefont {Su}}, \ and\ \bibinfo {author}
  {\bibfnamefont {J.-P.}\ \bibnamefont {Locquet}},\ }\href
  {http://dx.doi.org/10.1063/1.4866004} {\bibfield  {journal} {\bibinfo
  {journal} {Applied Physics Letters}\ }\textbf {\bibinfo {volume} {104}},\
  \bibinfo {pages} {071902} (\bibinfo {year} {2014})}\BibitemShut {NoStop}%
\bibitem [{\citenamefont {Gilbert}\ \emph {et~al.}(2017)\citenamefont
  {Gilbert}, \citenamefont {Ram{\'\i}rez}, \citenamefont {Saerbeck},
  \citenamefont {Trastoy}, \citenamefont {Schuller}, \citenamefont {Liu},\ and\
  \citenamefont {de~la Venta}}]{Gilbert_2017}%
  \BibitemOpen
  \bibfield  {author} {\bibinfo {author} {\bibfnamefont {D.~A.}\ \bibnamefont
  {Gilbert}}, \bibinfo {author} {\bibfnamefont {J.~G.}\ \bibnamefont
  {Ram{\'\i}rez}}, \bibinfo {author} {\bibfnamefont {T.}~\bibnamefont
  {Saerbeck}}, \bibinfo {author} {\bibfnamefont {J.}~\bibnamefont {Trastoy}},
  \bibinfo {author} {\bibfnamefont {I.~K.}\ \bibnamefont {Schuller}}, \bibinfo
  {author} {\bibfnamefont {K.}~\bibnamefont {Liu}}, \ and\ \bibinfo {author}
  {\bibfnamefont {J.}~\bibnamefont {de~la Venta}},\ }\href {\doibase
  10.1038/s41598-017-12690-z} {\bibfield  {journal} {\bibinfo  {journal}
  {Scientific Reports}\ }\textbf {\bibinfo {volume} {7}},\ \bibinfo {pages}
  {13471} (\bibinfo {year} {2017})}\BibitemShut {NoStop}%
\bibitem [{\citenamefont {de~la Venta}\ \emph {et~al.}(2014)\citenamefont
  {de~la Venta}, \citenamefont {Wang}, \citenamefont {Saerbeck}, \citenamefont
  {Ram{\'\i}rez}, \citenamefont {Valmianski},\ and\ \citenamefont
  {Schuller}}]{Venta.APL.2014}%
  \BibitemOpen
  \bibfield  {author} {\bibinfo {author} {\bibfnamefont {J.}~\bibnamefont
  {de~la Venta}}, \bibinfo {author} {\bibfnamefont {S.}~\bibnamefont {Wang}},
  \bibinfo {author} {\bibfnamefont {T.}~\bibnamefont {Saerbeck}}, \bibinfo
  {author} {\bibfnamefont {J.~G.}\ \bibnamefont {Ram{\'\i}rez}}, \bibinfo
  {author} {\bibfnamefont {I.}~\bibnamefont {Valmianski}}, \ and\ \bibinfo
  {author} {\bibfnamefont {I.~K.}\ \bibnamefont {Schuller}},\ }\href
  {http://scitation.aip.org/content/aip/journal/apl/104/6/10.1063/1.4865587}
  {\bibfield  {journal} {\bibinfo  {journal} {Applied Physics Letters}\
  }\textbf {\bibinfo {volume} {104}},\ \bibinfo {eid} {062410} (\bibinfo {year}
  {2014})}\BibitemShut {NoStop}%
\bibitem [{\citenamefont {Sass}\ \emph {et~al.}(2004)\citenamefont {Sass},
  \citenamefont {Tusche}, \citenamefont {Felsch}, \citenamefont {Quaas},
  \citenamefont {Weismann},\ and\ \citenamefont {Wenderoth}}]{Sass_2004}%
  \BibitemOpen
  \bibfield  {author} {\bibinfo {author} {\bibfnamefont {B.}~\bibnamefont
  {Sass}}, \bibinfo {author} {\bibfnamefont {C.}~\bibnamefont {Tusche}},
  \bibinfo {author} {\bibfnamefont {W.}~\bibnamefont {Felsch}}, \bibinfo
  {author} {\bibfnamefont {N.}~\bibnamefont {Quaas}}, \bibinfo {author}
  {\bibfnamefont {A.}~\bibnamefont {Weismann}}, \ and\ \bibinfo {author}
  {\bibfnamefont {M.}~\bibnamefont {Wenderoth}},\ }\href@noop {} {\bibfield
  {journal} {\bibinfo  {journal} {J. Phys.: Condens. Matter}\ }\textbf
  {\bibinfo {volume} {16}},\ \bibinfo {pages} {77} (\bibinfo {year}
  {2004})}\BibitemShut {NoStop}%
\bibitem [{\citenamefont {Ji}\ \emph {et~al.}(2012)\citenamefont {Ji},
  \citenamefont {Pan}, \citenamefont {Bi}, \citenamefont {Liang}, \citenamefont
  {Zhang}, \citenamefont {Zeng}, \citenamefont {Wen}, \citenamefont {Zhang},
  \citenamefont {Chen}, \citenamefont {Jia},\ and\ \citenamefont
  {Lin}}]{Ji.APL.2012}%
  \BibitemOpen
  \bibfield  {author} {\bibinfo {author} {\bibfnamefont {Y.~D.}\ \bibnamefont
  {Ji}}, \bibinfo {author} {\bibfnamefont {T.~S.}\ \bibnamefont {Pan}},
  \bibinfo {author} {\bibfnamefont {Z.}~\bibnamefont {Bi}}, \bibinfo {author}
  {\bibfnamefont {W.~Z.}\ \bibnamefont {Liang}}, \bibinfo {author}
  {\bibfnamefont {Y.}~\bibnamefont {Zhang}}, \bibinfo {author} {\bibfnamefont
  {H.~Z.}\ \bibnamefont {Zeng}}, \bibinfo {author} {\bibfnamefont {Q.~Y.}\
  \bibnamefont {Wen}}, \bibinfo {author} {\bibfnamefont {H.~W.}\ \bibnamefont
  {Zhang}}, \bibinfo {author} {\bibfnamefont {C.~L.}\ \bibnamefont {Chen}},
  \bibinfo {author} {\bibfnamefont {Q.~X.}\ \bibnamefont {Jia}}, \ and\
  \bibinfo {author} {\bibfnamefont {Y.}~\bibnamefont {Lin}},\ }\href
  {http://dx.doi.org/10.1063/1.4745843} {\bibfield  {journal} {\bibinfo
  {journal} {Applied Physics Letters}\ }\textbf {\bibinfo {volume} {101}},\
  \bibinfo {pages} {071902} (\bibinfo {year} {2012})}\BibitemShut {NoStop}%
\bibitem [{NBS(1983)}]{NBS_V2O3_bulk}%
  \BibitemOpen
  \href@noop {} {\bibfield  {journal} {\bibinfo  {journal} {National Bureau of
  Standards (US) Monograph}\ }\textbf {\bibinfo {volume} {25}},\ \bibinfo
  {pages} {108} (\bibinfo {year} {1983})}\BibitemShut {NoStop}%
\bibitem [{\citenamefont {Eckert}\ and\ \citenamefont
  {Bradt}(1973)}]{Eckert_JAP_1973}%
  \BibitemOpen
  \bibfield  {author} {\bibinfo {author} {\bibfnamefont {L.~J.}\ \bibnamefont
  {Eckert}}\ and\ \bibinfo {author} {\bibfnamefont {R.~C.}\ \bibnamefont
  {Bradt}},\ }\href {http://dx.doi.org/10.1063/1.1662787} {\bibfield  {journal}
  {\bibinfo  {journal} {Journal of Applied Physics}\ }\textbf {\bibinfo
  {volume} {44}},\ \bibinfo {pages} {3470} (\bibinfo {year}
  {1973})}\BibitemShut {NoStop}%
\bibitem [{\citenamefont {Munro}(1997)}]{Munro_1997}%
  \BibitemOpen
  \bibfield  {author} {\bibinfo {author} {\bibfnamefont {R.~G.}\ \bibnamefont
  {Munro}},\ }\href@noop {} {\bibfield  {journal} {\bibinfo  {journal} {J. Am.
  Ceram. Soc.}\ }\textbf {\bibinfo {volume} {80}},\ \bibinfo {pages} {1919}
  (\bibinfo {year} {1997})}\BibitemShut {NoStop}%
\bibitem [{\citenamefont {Gao}\ \emph {et~al.}(2002)\citenamefont {Gao},
  \citenamefont {Lu},\ and\ \citenamefont {Suo}}]{Gao_2002}%
  \BibitemOpen
  \bibfield  {author} {\bibinfo {author} {\bibfnamefont {Y.}~\bibnamefont
  {Gao}}, \bibinfo {author} {\bibfnamefont {W.}~\bibnamefont {Lu}}, \ and\
  \bibinfo {author} {\bibfnamefont {Z.}~\bibnamefont {Suo}},\ }\href {\doibase
  https://doi.org/10.1016/S1359-6454(02)00056-3} {\bibfield  {journal}
  {\bibinfo  {journal} {Acta Materialia}\ }\textbf {\bibinfo {volume} {50}},\
  \bibinfo {pages} {2297 } (\bibinfo {year} {2002})}\BibitemShut {NoStop}%
\bibitem [{\citenamefont {Bratkovsky}\ \emph {et~al.}(1994)\citenamefont
  {Bratkovsky}, \citenamefont {Marais}, \citenamefont {Heine},\ and\
  \citenamefont {Salje}}]{Bratkovsky_1994}%
  \BibitemOpen
  \bibfield  {author} {\bibinfo {author} {\bibfnamefont {A.~M.}\ \bibnamefont
  {Bratkovsky}}, \bibinfo {author} {\bibfnamefont {S.~C.}\ \bibnamefont
  {Marais}}, \bibinfo {author} {\bibfnamefont {V.}~\bibnamefont {Heine}}, \
  and\ \bibinfo {author} {\bibfnamefont {E.~K.~H.}\ \bibnamefont {Salje}},\
  }\href {http://stacks.iop.org/0953-8984/6/i=20/a=008} {\bibfield  {journal}
  {\bibinfo  {journal} {Journal of Physics: Condensed Matter}\ }\textbf
  {\bibinfo {volume} {6}},\ \bibinfo {pages} {3679} (\bibinfo {year}
  {1994})}\BibitemShut {NoStop}%
\bibitem [{\citenamefont {Tokura}\ and\ \citenamefont
  {Nagaosa}(2000)}]{Tokura_Science_2000}%
  \BibitemOpen
  \bibfield  {author} {\bibinfo {author} {\bibfnamefont {Y.}~\bibnamefont
  {Tokura}}\ and\ \bibinfo {author} {\bibfnamefont {N.}~\bibnamefont
  {Nagaosa}},\ }\href {\doibase 10.1126/science.288.5465.462} {\bibfield
  {journal} {\bibinfo  {journal} {Science}\ }\textbf {\bibinfo {volume}
  {288}},\ \bibinfo {pages} {462} (\bibinfo {year} {2000})}\BibitemShut
  {NoStop}%
\bibitem [{\citenamefont {He}\ and\ \citenamefont
  {Millis}(2016)}]{He_PRB_2016}%
  \BibitemOpen
  \bibfield  {author} {\bibinfo {author} {\bibfnamefont {Z.}~\bibnamefont
  {He}}\ and\ \bibinfo {author} {\bibfnamefont {A.~J.}\ \bibnamefont
  {Millis}},\ }\href {\doibase 10.1103/PhysRevB.93.115126} {\bibfield
  {journal} {\bibinfo  {journal} {Phys. Rev. B}\ }\textbf {\bibinfo {volume}
  {93}},\ \bibinfo {pages} {115126} (\bibinfo {year} {2016})}\BibitemShut
  {NoStop}%
\bibitem [{\citenamefont {Morrison}\ \emph {et~al.}(2014)\citenamefont
  {Morrison}, \citenamefont {Chatelain}, \citenamefont {Tiwari}, \citenamefont
  {Hendaoui}, \citenamefont {Bruh{\'a}cs}, \citenamefont {Chaker},\ and\
  \citenamefont {Siwick}}]{Morrison_Science_2014}%
  \BibitemOpen
  \bibfield  {author} {\bibinfo {author} {\bibfnamefont {V.~R.}\ \bibnamefont
  {Morrison}}, \bibinfo {author} {\bibfnamefont {R.~P.}\ \bibnamefont
  {Chatelain}}, \bibinfo {author} {\bibfnamefont {K.~L.}\ \bibnamefont
  {Tiwari}}, \bibinfo {author} {\bibfnamefont {A.}~\bibnamefont {Hendaoui}},
  \bibinfo {author} {\bibfnamefont {A.}~\bibnamefont {Bruh{\'a}cs}}, \bibinfo
  {author} {\bibfnamefont {M.}~\bibnamefont {Chaker}}, \ and\ \bibinfo {author}
  {\bibfnamefont {B.~J.}\ \bibnamefont {Siwick}},\ }\href {\doibase
  10.1126/science.1253779} {\bibfield  {journal} {\bibinfo  {journal}
  {Science}\ }\textbf {\bibinfo {volume} {346}},\ \bibinfo {pages} {445}
  (\bibinfo {year} {2014})}\BibitemShut {NoStop}%
\bibitem [{\citenamefont {Aetukuri}\ \emph {et~al.}(2013)\citenamefont
  {Aetukuri}, \citenamefont {Gray}, \citenamefont {Drouard}, \citenamefont
  {Cossale}, \citenamefont {Gao}, \citenamefont {Reid}, \citenamefont
  {Kukreja}, \citenamefont {Ohldag}, \citenamefont {Jenkins}, \citenamefont
  {Arenholz}, \citenamefont {Roche}, \citenamefont {D{\"u}rr}, \citenamefont
  {Samant},\ and\ \citenamefont {Parkin}}]{Aetukuri_NatPhys_2013}%
  \BibitemOpen
  \bibfield  {author} {\bibinfo {author} {\bibfnamefont {N.~B.}\ \bibnamefont
  {Aetukuri}}, \bibinfo {author} {\bibfnamefont {A.~X.}\ \bibnamefont {Gray}},
  \bibinfo {author} {\bibfnamefont {M.}~\bibnamefont {Drouard}}, \bibinfo
  {author} {\bibfnamefont {M.}~\bibnamefont {Cossale}}, \bibinfo {author}
  {\bibfnamefont {L.}~\bibnamefont {Gao}}, \bibinfo {author} {\bibfnamefont
  {A.~H.}\ \bibnamefont {Reid}}, \bibinfo {author} {\bibfnamefont
  {R.}~\bibnamefont {Kukreja}}, \bibinfo {author} {\bibfnamefont
  {H.}~\bibnamefont {Ohldag}}, \bibinfo {author} {\bibfnamefont {C.~A.}\
  \bibnamefont {Jenkins}}, \bibinfo {author} {\bibfnamefont {E.}~\bibnamefont
  {Arenholz}}, \bibinfo {author} {\bibfnamefont {K.~P.}\ \bibnamefont {Roche}},
  \bibinfo {author} {\bibfnamefont {H.~A.}\ \bibnamefont {D{\"u}rr}}, \bibinfo
  {author} {\bibfnamefont {M.~G.}\ \bibnamefont {Samant}}, \ and\ \bibinfo
  {author} {\bibfnamefont {S.~S.~P.}\ \bibnamefont {Parkin}},\ }\href
  {http://dx.doi.org/10.1038/nphys2733} {\bibfield  {journal} {\bibinfo
  {journal} {Nature Physics}\ }\textbf {\bibinfo {volume} {9}},\ \bibinfo
  {pages} {661 EP } (\bibinfo {year} {2013})}\BibitemShut {NoStop}%
\bibitem [{\citenamefont {Verma}\ and\ \citenamefont {Ram}(2016)}]{Verma_2016}%
  \BibitemOpen
  \bibfield  {author} {\bibinfo {author} {\bibfnamefont {M.}~\bibnamefont
  {Verma}}\ and\ \bibinfo {author} {\bibfnamefont {K.}~\bibnamefont {Ram}},\
  }\href {\doibase 10.1063/1.4948750} {\bibfield  {journal} {\bibinfo
  {journal} {AIP Advances}\ }\textbf {\bibinfo {volume} {6}},\ \bibinfo {pages}
  {055302} (\bibinfo {year} {2016})}\BibitemShut {NoStop}%
\end{thebibliography}

%

\end{document}